\documentclass[aps, prl, reprint, showpacs, twocolumn, 10pt, groupedaddress]{revtex4-2}

\usepackage{graphicx, amsmath, dsfont, dcolumn, bm, times, color, braket, amssymb, multirow}

\begin{document}

\bibliographystyle{apsrev4-2}

\title{Signature of PT-symmetric non-Hermitian superconductivity in angle-resolved photoelectron fluctuation spectroscopy}

\author{Viktoriia Kornich}
\affiliation{Institut f\"ur Theoretische Physik and Astrophysik, Universit\"at W\"urzburg, 97074 W\"urzburg, Germany }
\author{Bj\"orn Trauzettel}
\affiliation{Institut f\"ur Theoretische Physik and Astrophysik, Universit\"at W\"urzburg, 97074 W\"urzburg, Germany }

\date{\today}

\begin{abstract}
We show theoretically that the measurement of a $\mathcal{PT}$-symmetric non-Hermitian superconductor by angle-resolved photoelectron fluctuation spectroscopy (ARPFS) provides a decisive signature. In fact, the signal is negative in a single-band case in contrast to the ARPFS signal of Hermitian superconductors. We suggest that the negative fluctuations can be explained by a remarkable pairing phenomenon: If the interaction between electrons in this $\mathcal{PT}$-symmetric non-Hermitian superconductor is attractive then the interaction between holes (i.~e.~missing electrons) is repulsive and vice versa. This difference in the sign of the interactions gives rise to negative cross correlations. Here, we propose how such electron-electron interaction can occur due to spatiotemporal modulation of the material. We also discuss the observability of this signature in multi-band systems.
\end{abstract}

\maketitle

\let\oldvec\vec
\renewcommand{\vec}[1]{\ensuremath{\boldsymbol{#1}}}
{\it Introduction.--}
Exotic states of matter allow for significant advances in fundamental science and sometimes even in technology. In this sense, the physics of non-Hermitian systems has acquired a lot of attention and evolved into a rapidly developing research field \cite{ashida:advphys20, kawabata:prx19}. Among these systems, $\mathcal{PT}$-symmetric structures \cite{bender:prl98, bender:rpp07} have received particular interest due to their multiple applications in optics and synthetic materials. Here, $\mathcal{P}$ refers to parity and $\mathcal{T}$ to time-reversal. It has been predicted that adding optical loss and gain to Hermitian systems allows to form $\mathcal{PT}$-symmetric systems in the laboratory \cite{makris:prl08} that exhibit exciting physics, for instance, loss-induced optical transparency \cite{guo:prl09} and reversing the pump dependence of a laser \cite{brandstetter:natcom14}. There are many other interesting effects and proposed applications based on $\mathcal{PT}$-symmetry of optical systems \cite{ozdemir:natmater19}, e.g. visualization of exceptional points in $\mathcal{PT}$-symmetric directional couplers \cite{klaiman:prl08}, laser absorbers \cite{longhi:pra10}, unidirectional invisibility of media \cite{regensburger:nature12}, and selective mode lasers \cite{feng:science14, hodaei:science14}.

$\mathcal{PT}$-symmetric superconductors have not been studied as extensively as optical $\mathcal{PT}$-symmetric systems. However, a number of recent papers report particular properties related to $\mathcal{PT}$-symmetric superconductors and corresponding hybrid structures. For example, $\mathcal{PT}$-symmetric superconductors have been studied in relation to Majorana fermions with theoretical works showing unusual anticommutation relations \cite{kawabata:prb18} or dragging of mobile Majorana fermions \cite{yang:prr20}. It has been reported in Ref. \cite{chtchelkatchev:prl12} that $\mathcal{PT}$-symmetry stabilizes superconductivity near the phase transition in a 1D system. In Refs.~\cite{zhao:prl16, yuce:scirep18}, the authors have discussed the theory of $\mathcal{PT}$-invariant topological metals, semimetals, and nodal superconductors from a more general perspective. Moreover, the superconducting $\mathcal{PT}$-symmetric phase transition in metasurfaces has experimentally been investigated in Ref.~\cite{wang:apl17}. Non-Hermitian superconductors with $\mathcal{PT}$-symmetric Cooper pairing have been theoretically studied in Ref.~\cite{ghatak:prb18}, where Dzyaloshinksii-Moriya interaction in combination with an external bath or the imbalance between electron-electron and hole-hole pairs has been suggested as possible origin of $\mathcal{PT}$-symmetric pairing. The theory of non-Hermitian fermionic superfluidity with a complex-valued interaction have been discussed in Ref.~\cite{yamamoto:prl19}.

Taking into account the variety of possible applications of non-Hermitian $\mathcal{PT}$-symmetric superconductors, it is important to develop detection schemes that allow for their unique identification. It has been shown theoretically that angle-resolved photoelectron fluctuation spectroscopy (ARPFS) can be used for measuring the anomalous Green's function characterizing the superconducting state, in Hermitian systems \cite{stahl:prb19}. This theory has been further developed to propose the direct detection of the ``order parameter'' of odd-frequency superconductivity \cite{kornich:prr21}.

 \begin{figure}[tb]
 	\begin{center}
 		\includegraphics[width=\linewidth]{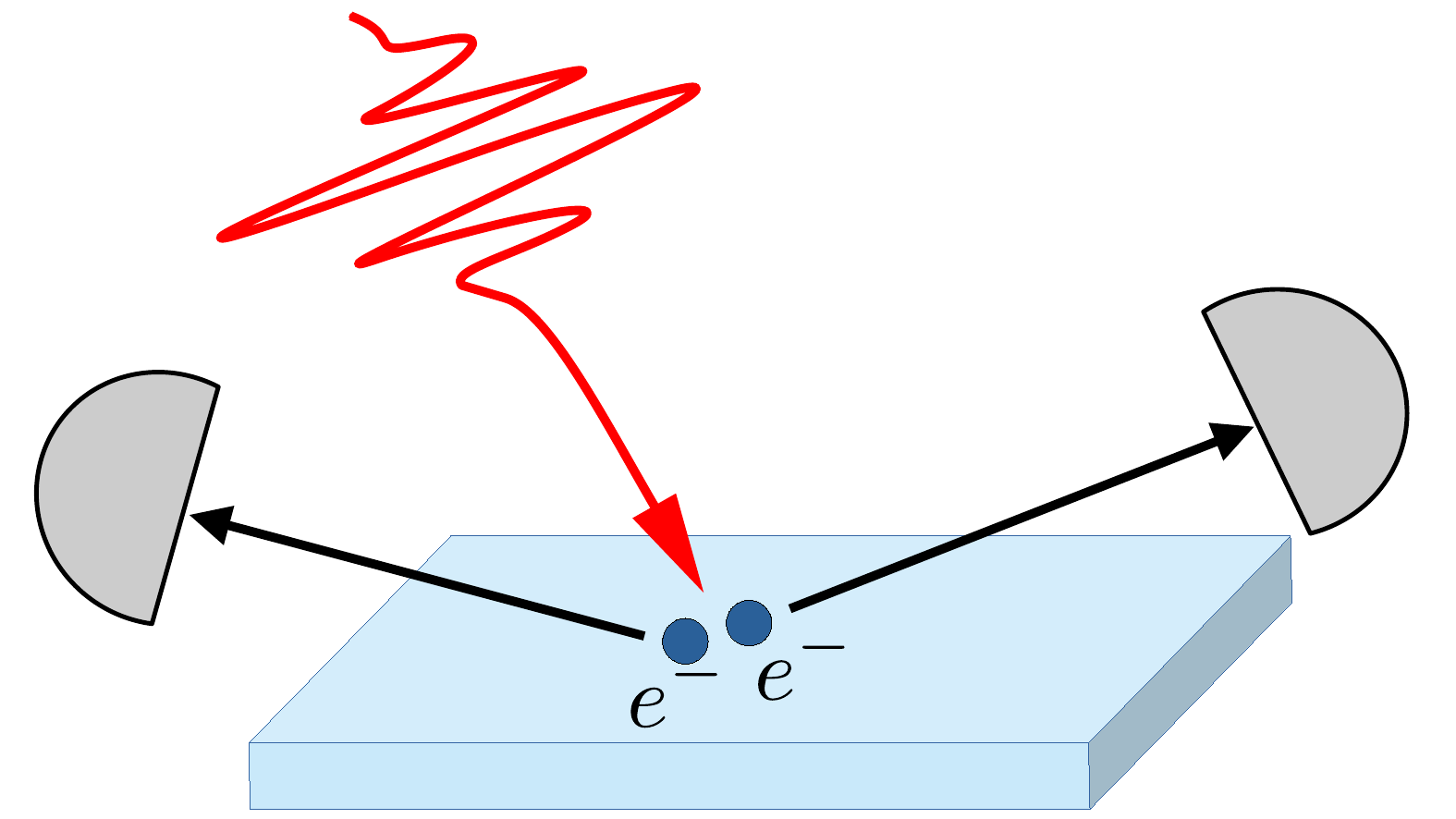}
 		\caption{Schematic of the ARPFS setup: An electromagnetic pulse is hitting onto the $\mathcal{PT}$-symmetric superconductor. Electrons are emitted from the substrate due to interaction with photons from the pulse and are registered by two detectors located at opposite sides of the sample. The aim of the detection scheme is to identify correlated electrons that formed a Cooper pair in the sample before they have left it.}
 		\label{fig:setup}
 	\end{center}
 \end{figure}

In this work, we suggest to detect non-Hermitian $\mathcal{PT}$-symmetric superconductors via ARPFS. We prove that the ARPFS signal for a single-band $\mathcal{PT}$-symmetric non-Hermitian superconductor is negative. (Note that the corresponding signal for a Hermitian superconductor has to be positive.) This remarkable result implies that it is unfavourable for such a system to photoemit a correlated pair of electrons. We suggest that this phenomenon can be explained by the asymmetric electron-electron interaction potential that effectively leads to attractive interaction of electrons and repulsive interaction of corresponding holes or vice versa.

{\it $\mathcal{PT}$-symmetric non-Hermitian superconductor.--} The BCS-type (mean-field) Hamiltonian under consideration is
\begin{eqnarray}
\label{eq:HamiltonianSC}
\nonumber H&=&\sum_{\bf k}\Psi_{\bf k}^\dagger\mathcal{H}_{\bf k}\Psi_{\bf k}=\\ &=&\sum_{\bf k} \Psi_{\bf k}^\dagger(\varepsilon_k \tau_z+i\sigma_y[\Delta_{\bf k}\tau_+-\bar{\Delta}_{\bf k}\tau_-])\Psi_{\bf k},
\end{eqnarray}
where $\Psi_{\bf k}^\dagger=\{c^\dagger_{{\bf k},\uparrow},c^\dagger_{{\bf k},\downarrow}, c_{-{\bf k},\uparrow}, c_{-{\bf k},\downarrow}\}$ and $c_{{\bf k},\sigma}$ is the annihilation operator of an electron with momentum ${\bf k}$ and spin $\sigma$; the spin quantization axis is $z$; $\tau_z$ and $\sigma_y$ are the corresponding Pauli matrices in particle-hole and spin spaces, respectively, with $\tau_\pm=(\tau_x\pm i\tau_y)/2$; $\varepsilon_k$ is the spectrum of electrons (in the absence of superconductivity) and we assume $\varepsilon_k$ to be real; $\Delta_{\bf k}$ and $\bar{\Delta}_{\bf k}$ denote mean-field pairing potentials. In the Hermitian case, the following equality holds $\bar{\Delta}_{\bf k}=\Delta^\dagger_{\bf k}$. We now explain the conditions that apply to $\Delta_{\bf k}$ and $\bar{\Delta}_{\bf k}$ based on conservation of $\mathcal{PT}$ symmetry. If the Hamiltonian is invariant with respect to $\mathcal{PT}$ symmetry this implies that $(\mathcal{PT})\mathcal{H}_{\bf k}(\mathcal{PT})^{-1}=\mathcal{H}_{\bf k}$. The time-reversal operator $\mathcal{T}$ is defined as $\mathcal{T}=-i\sigma_y\mathcal{K}$, where $\mathcal{K}$ is the operator of complex conjugation in real space (thus inverting the sign of momentum). The parity (or space inversion/reflection operator) $\mathcal{P}$ inverts the sign of the momentum. Thus, the Hamiltonian (\ref{eq:HamiltonianSC}) is $\mathcal{PT}$-symmetric, if $\mathcal{H}_{\bf k}=\mathcal{H}_{\bf k}^*$. This is possible for real mean fields $\Delta_{\bf k}$ and $\bar{\Delta}_{\bf k}$. Assuming that the superconductor is not pumped with electrons or holes, we take $|\Delta_{\bf k}|=|\bar{\Delta}_{\bf k}|$.

Now, we introduce and discuss non-Hermitian characteristics of the Hamiltonian (\ref{eq:HamiltonianSC}). Non-Hermitian theory implies that $\Delta_{\bf k}^\dagger\neq \bar{\Delta}_{\bf k}$. Taking into account all the above conditions, namely $|\Delta_{\bf k}|=|\bar{\Delta}_{\bf k}|$ and $\Delta_{\bf k}, \bar{\Delta}_{\bf k}\in \Re$, the Hamiltonian (\ref{eq:HamiltonianSC}) can be non-Hermitian only if we define the mean fields to be anti-Hermitian, i.e. $\Delta_{\bf k}^\dagger=-\bar{\Delta}_{\bf k}$, as in Ref. \cite{ghatak:prb18}. Then the excitation energies of $\mathcal{H}_{\bf k}$ are $\epsilon_{\bf k}=\pm\sqrt{\varepsilon_k^2-|\Delta_{\bf k}|^2}$. Thus, for $|\varepsilon_k|>|\Delta_{\bf k}|$, the excitation spectrum is real, even though the underlying BdG Hamiltonian is non-Hermitian. For $|\varepsilon_k|<|\Delta_{\bf k}|$, instead, the eigenvalues are complex and appear in complex conjugate pairs. In our work, we focus on the regime $|\varepsilon_k|>|\Delta_{\bf k}|$ implying a real spectrum. From there it follows that $\mathcal{H}_{\bf k}$ can be understood as pseudo-Hermitian \cite{mostafazadeh:jmp02}, meaning $\mathcal{H}_{\bf k}^\dagger=\eta \mathcal{H}_{\bf k}\eta^{-1}$ with $\eta$ being a Hermitian and invertible linear operator. This relation implies that the orthonormal scalar product must be defined for the $\eta$-product of left and right eigen states: $\langle\Phi_L^\alpha|\Phi_R^\beta\rangle=\langle\Phi_R^\alpha|\eta|\Phi_R^\beta\rangle=\delta_{\alpha\beta}$.

{\it Non-Hermitian theory of ARPFS.--}
The theory of ARPFS in the Hermitian case has been carefully described before, see e.g. Ref. \cite{stahl:prb19}. Here, we shortly present the important equations emphasizing the points related to the non-Hermitian aspects of the derivation. The Hamiltonian of light-matter interaction is given by
\begin{eqnarray}
H=\sum_{k,p,\sigma,\sigma'}S^*(t)e^{i\Omega t}M^{\sigma,\sigma'}_{{\bf k},{\bf p}}f^\dagger_{{\bf p}, \sigma'}c_{{\bf k},\sigma}+H.c.,
\end{eqnarray}
where $M^{\sigma,\sigma'}_{{\bf k},{\bf p}}$ are the matrix elements for emission and $S(t)$ is the temporal envelope of the probe with frequency $\Omega$ centered around $t$. The operator $f^\dagger_{{\bf p},\sigma'}$ is the creation operator for emitted electrons. The Hamiltonian of emitted electrons has the simple form $H_f=\sum_{{\bf p},\sigma}E_pf^\dagger_{{\bf p},\sigma}f_{{\bf p},\sigma}=\sum_{{\bf p},\sigma}E_pn_{{\bf p},\sigma}$. The total population of emitted electrons can be written as
\begin{eqnarray}
\label{eq:population}
I^{(1)}_{{\bf p},\sigma}=\langle n_{{\bf p},\sigma}\rangle=\langle \mathcal{S}_L n_{{\bf p},\sigma}\mathcal{S}_R\rangle_0,
\end{eqnarray}
where the index $0$ denotes averaging with respect to the ground state and its $\eta$-pseudo-Hermitian adjoint $\langle...\rangle_0=\langle0|\eta...|0\rangle$, see also \cite{SM}. The evolution operator for the right state vector is $\mathcal{S}_{R}=T\exp{\left[- i\int_{-\infty}^{\infty}d\tau H(\tau) \right]}$ and for the left state vector is $\mathcal{S}_{L}=\bar{T}\exp{\left[ i\int_{-\infty}^{\infty}d\tau H(\tau) \right]}$. We can see that with this choice of left and right basis states, $\mathcal{S}_R$ coincides with the known expression for the $\mathcal{S}$-matrix and $\mathcal{S}_L$ coincides with $\mathcal{S}^\dagger$. Here, $T$ and $\bar{T}$ denote time and anti-time ordering, respectively. 

The statistical correlations of photoemission events read
\begin{eqnarray}
I^{(2)}_{{\bf p},\sigma;{\bf p'},\sigma'}=\langle n_{{\bf p},\sigma}n_{{\bf p}',\sigma'}\rangle,
\end{eqnarray}
where the average $\langle ...\rangle$ is defined in the same way as in Eq.~(\ref{eq:population}). In order to calculate $I^{(2)}_{{\bf p},\sigma;{\bf p'},\sigma'}$, we expand $\mathcal{S}_L$ and $\mathcal{S}_R$ up to second order assuming weak light-matter interaction. This implies that we need to average eight $f$-operators and four $c$-operators. The emitted electrons refer to a quadratic Hermitian Hamiltonian. Thus, we average them using Wick's theorem. The $c$-operators can also be averaged using Wick's theorem, if we carefully define the left and the right basis \cite{herviou:scipost19}. Thus, we can decouple the two-point Green's function $G^{{\bf k}_2,{\bf k}_1,{\bf k}_1',{\bf k}_2'}_{\sigma_2,\sigma_1,\sigma_1',\sigma_2'}(\tau_2,\tau_1,\tau_1',\tau_2')=\langle\bar{T}[c^\dagger_{{\bf k}_2,\sigma_2}(\tau_2)c^\dagger_{{\bf k}_1,\sigma_1}(\tau_1)]T[c_{{\bf k}_1',\sigma_1'}(\tau_1')c_{{\bf k}_2',\sigma_2'}(\tau_2')]\rangle_0$ into correlations of pairs of operators of the form $\langle c^\dagger c\rangle_0$, $\langle c^\dagger c^\dagger\rangle_0$, and $\langle cc\rangle_0$.
The fluctuations of the correlations of photoelectrons are defined as
\begin{eqnarray}
\Delta I_{{\bf p},\sigma;{\bf p'},\sigma'}=I^{(2)}_{{\bf p},\sigma;{\bf p'},\sigma'}-I^{(1)}_{{\bf p},\sigma}I^{(1)}_{{\bf p'},\sigma'}.
\end{eqnarray}
In the case ${\bf p'}=-{\bf p}$ (corresponding to the setup shown in Fig. \ref{fig:setup}) and for the Hamiltonian defined in Eq. (\ref{eq:HamiltonianSC}), the non-zero terms of the type $\langle c^\dagger c\rangle_0$ from $I^{(2)}_{{\bf p},\sigma;{\bf p'},\sigma'}$ cancel with the ones from $I^{(1)}_{{\bf p},\sigma}I^{(1)}_{{\bf p'},\sigma'}$. Thus, only terms depending on the anomalous Green's functions, defined as $F_{\sigma',\sigma}^{-{\bf p},{\bf p}}(\tau_1,\tau_2)=\langle T[c_{-{\bf p},\sigma'}(\tau_1)c_{{\bf p},\sigma}(\tau_2)]\rangle_0$ and $\bar{F}_{\sigma,\sigma'}^{{\bf p},-{\bf p}}(\tau_2,\tau_1)=\langle \bar{T}[c^\dagger_{{\bf p},\sigma}(\tau_2)c^\dagger_{-{\bf p},\sigma'}(\tau_1)]\rangle_0$, remain. We note that in general $\bar{F}\neq F^\dagger$ for a non-Hermitian system. If we employ the simplified matrix element form $M_{{\bf k},{\bf p}}^{\sigma,\sigma'}=M_0\delta_{{\bf k},{\bf p}}\delta_{\sigma,\sigma'}$, a usual approximation in the theory of angle-resolved photoemission spectroscopy, we obtain
\begin{widetext}
\begin{eqnarray}
\label{eq:DeltaIFull}
\Delta I_{{\bf p},\sigma;-{\bf p},\sigma'}=M_0^4\int_{-\infty}^\infty d\tau_1d\tau_2d\tau_1'd\tau_2' S(\tau_1)S(\tau_2)S^*(\tau_1')S^*(\tau_2')e^{i(\Omega+E_p) (\tau_1'+\tau_2'-\tau_1-\tau_2)}\bar{F}_{\sigma,\sigma'}^{{\bf p},-{\bf p}}(\tau_2,\tau_1)F_{\sigma',\sigma}^{-{\bf p},{\bf p}}(\tau_1',\tau_2').
\end{eqnarray}
\end{widetext}
Notice, that Eq. (\ref{eq:DeltaIFull}) contains cross-correlations of the anomalous Green's functions of photoemitted electrons. Positive values of $\Delta I_{{\bf p},\sigma;-{\bf p},\sigma'}$ refer to correlations, while negative values refer to anti-correlations.

{\it ARPFS signal for $\mathcal{PT}$-symmetric non-Hermitian superconductors.--} Let us apply this result to the case of a $\mathcal{PT}$-symmetric non-Hermitian superconductor, as specified above.
The Green's function $\mathcal{G}=[1-H]^{-1}$ can be written as
\begin{eqnarray}
\label{eq:G}
\mathcal{G}=\begin{pmatrix}G & F\\ \bar{F} & \bar{G}\end{pmatrix}=\frac{\omega+\varepsilon_k\tau_z+i\sigma_y(\tau_+\Delta_{\bf k}-\tau_-\bar{\Delta}_{\bf k})}{\omega^2-\varepsilon_k^2-\Delta_{\bf k}\bar{\Delta}_{\bf k}{\color{white}\overrightarrow{1}}}.
\end{eqnarray}
Taking into account that $\bar{\Delta}_{\bf k}=-{\Delta}_{\bf k}$ with $\Delta_{\bf k}\in \Re$, we obtain, for instance, for $\sigma=\uparrow$ and $\sigma'=\downarrow$,
\begin{eqnarray}
\label{eq:F}
F_{\downarrow, \uparrow}^{-{\bf k},{\bf k}}(\omega)=-\bar{F}_{\uparrow,\downarrow}^{{\bf k},-{\bf k}}(\omega)=\frac{-\Delta_{\bf k}}{\omega^2-\varepsilon_k^2+\Delta_{\bf k}^2}.
\end{eqnarray}
In order to apply these anomalous Green's functions to Eq. (\ref{eq:DeltaIFull}), we need to perform Fourier transformation of Eq. (\ref{eq:F}). Building a contour of half-circle shape with the radius $R\rightarrow\infty$ and employing residues, see \cite{SM}, we obtain for the anomalous Green's functions in the time domain
\begin{eqnarray}
\bar{F}_{\uparrow,\downarrow}^{{\bf k},-{\bf k}}(t)&=&-\frac{i\Delta_{\bf k}e^{-it\sqrt{\varepsilon_k^2-\Delta_{\bf k}^2}}}{2\sqrt{\varepsilon_k^2-\Delta_{\bf k}^2}},\\
F_{\downarrow,\uparrow}^{-{\bf k},{\bf k}}(t)&=&-\frac{i\Delta_{\bf k}e^{it\sqrt{\varepsilon_k^2-\Delta_{\bf k}^2}}}{2\sqrt{\varepsilon_k^2-\Delta_{\bf k}^2}}.
\end{eqnarray}
This directly illustrates the fact that $\bar{F}_{\uparrow,\downarrow}^{{\bf k},-{\bf k}}(t)\neq (F_{\downarrow,\uparrow}^{-{\bf k},{\bf k}}(t))^\dagger$, and thus Eq.~(\ref{eq:DeltaIFull}) does not simplify to the absolute value squared of an integral over only one anomalous Green's function as in case of Hermitian superconductors \cite{stahl:prb19}. This implies that the ARPFS signal can be negative.

Next, we apply this result to Eq. (\ref{eq:DeltaIFull}). For simplicity, we assume delta-function-shaped pulses, $S=S_0\delta(t-t_0)$. Physically, this assumption means that the temporal width of the pulse is the shortest time scale under consideration. Thus, in corresponding experiments, we might need to take into account the relation of the frequency width of the pulse and the band structure of the sample in order to avoid photoemission from different bands. 
As an example, in order to emit electrons from a single band by a pulse of duration 30 fs, we need a band separation of at least 20 meV, when the probed momenta are near the Fermi momentum. This criterion is fulfilled in a variety of superconductors, e.g. Sr$_2$RuO$_4$ \cite{tamai:prx19} and iron pnictides \cite{nica:qm17}.

In our example of the single-band superconductor, the signal is given by
\begin{eqnarray}
\label{eq:NegativeFl}
\Delta I_{{\bf p},\uparrow;-{\bf p},\downarrow}=-\frac{M_0^4S_0^4}{4}\frac{\Delta_{\bf p}^2}{\varepsilon_p^2-\Delta_{\bf p}^2}.
\end{eqnarray}
Evidently, we obtain the particular result that the photoelectron fluctuations are negative in the regime under consideration, i.e. $|\varepsilon_p|>\Delta_{\bf p}$. In such experiments, the momentum ${\bf p}$ can be chosen. Therefore, we can always choose a large enough ${\bf p}$ to work in the desired regime. We underline that such signal appears due to $\bar{F}_{\uparrow,\downarrow}^{{\bf k},-{\bf k}}(t)=-(F_{\downarrow,\uparrow}^{-{\bf k},{\bf k}}(t))^\dagger$. Note that it can in principle be obtained in a non-$\mathcal{PT}$-symmetric case too. The main requirement is that the Hamiltonian of the type of Eq.~(\ref{eq:HamiltonianSC}) is non-Hermitian.

Let us suggest a physical explanation of Eq. (\ref{eq:NegativeFl}) for the non-Hermitian $\mathcal{PT}$-symmetric superconductor as defined above. We first draw the analogy to the case of two electrons placed at ${\bf r}_1$ and ${\bf r}_2$ interacting repulsively via Coulomb interaction. Their joint density $\rho({\bf r}_1,{\bf r}_2)$ is then smaller than the product of the independent densities $\rho({\bf r}_1)$ and $\rho({\bf r}_2)$, at small distances $|{\bf r}_1-{\bf r}_2|$, implying $\rho({\bf r}_1,{\bf r}_2)-\rho({\bf r}_1)\rho({\bf r}_2)<0$. In a notation related to the previous analysis, we can write this inequality as $\langle n_{{\bf p},\uparrow}n_{-{\bf p},\downarrow}\rangle-\langle n_{{\bf p},\uparrow}\rangle\langle n_{-{\bf p},\downarrow}\rangle<0$. In analogy to the ARPFS case discussed above, this means that after an electron with momentum ${\bf p}$ and spin $\uparrow$ has been emitted, the emission of an electron previously correlated with the first one with momentum $-{\bf p}$ and spin $\downarrow$ is suppressed. This suppression happens because, unlike in the Hermitian case, the hole mean field has the opposite sign to the electron mean-field. This implies that if electrons interact attractively, holes interact repulsively, and vice versa.

This surprising result can be formally understood as follows. The electron-electron interaction is described as
\begin{eqnarray}
\label{eq:Uel}
U=\sum_{{\bf p},{\bf q}}\psi^\dagger_{{\bf p}+{\bf q},\uparrow}\psi^\dagger_{-{\bf p}-{\bf q},\downarrow}V({\bf q})\psi_{-{\bf p},\downarrow}\psi_{{\bf p},\uparrow}.
\end{eqnarray}
It can be transformed into the corresponding hole interaction using fermionic commutation relations and relabelling the momentum indices as
\begin{eqnarray}
\label{eq:Uh}
U=\sum_{{\bf p},{\bf q}}\psi_{-{\bf p}-{\bf q},\downarrow}\psi_{{\bf p}+{\bf q},\uparrow}V(-{\bf q})\psi^\dagger_{{\bf p},\uparrow}\psi^\dagger_{-{\bf p},\downarrow}.
\end{eqnarray}
From Eqs. (\ref{eq:Uel}) and (\ref{eq:Uh}), we obtain expressions for the corresponding mean fields
\begin{eqnarray}
\bar{\Delta}_{\bf k}=\sum_{\bf p}V({\bf p}-{\bf k})\langle\psi^\dagger_{\bf p,\uparrow}\psi^\dagger_{-{\bf p},\downarrow}\rangle,\\
\Delta_{\bf k}=\sum_{\bf p}V({\bf k}-{\bf p})\langle\psi_{-{\bf p},\downarrow}\psi_{{\bf p},\uparrow}\rangle.
\end{eqnarray}
If the interaction potential $V({\bf q})$ is real and there is a balance between electrons and holes $\langle\psi^\dagger_{\bf p,\uparrow}\psi^\dagger_{-{\bf p},\downarrow}\rangle=\langle\psi_{-{\bf p},\downarrow}\psi_{{\bf p},\uparrow}\rangle$, the interaction potential $V({\bf q})$ must be odd in order to fulfil $\Delta_{\bf k}=-\bar{\Delta}_{\bf k}$. Eqs. (\ref{eq:Uel}) and (\ref{eq:Uh}) imply that electrons attract each other whereas holes repel each other (or vice versa). If a Cooper pair of electrons is photoemitted, a pair of holes is left behind. Since the holes repel each other this process costs energy. Hence, the photoemission of the two electrons that form the Cooper pair is suppressed.

Odd interaction potentials are typically related to certain asymmetries in the system, for instance, non-reciprocity or asymmetry of the spectrum. Non-reciprocity is characteristic to $\mathcal{PT}$-symmetric systems, e.g. non-reciprocal light transmission \cite{ramezani:pra10, rueter:natphys10, peng:nature14}, non-reciprocal bands in diatomic plasmonic chains \cite{ling:prb15}, or the asymmetric spectrum of a $\mathcal{PT}$-symmetric superconductor with two bands \cite{kanasugi:arXiv21}.

{\it Possible mechanism of electron-electron interaction in a $\mathcal{PT}$-symmetric non-Hermitian superconductor.--} We propose that the asymmetry leading to an odd interaction potential can also be achieved via external effects, such as the spatiotemporal modulation of material properties inducing non-reciprocity of the elastic wave propagation \cite{trainiti:njp16,zanjani:scirep15,zanjani:apl14,rasmussen:jap21}, that can modify electron-electron interaction correspondingly. We suggest to consider the following situation: Two phonon bands, $\omega_{1,{\bf q}}$ and $\omega_{2,{\bf q}}$, interact with each other strongly with an off-diagonal element $\delta_{\omega,{\bf q}}$ and due to non-reciprocity $\delta_{\pm\omega,{\bf q}}\gg \delta_{\pm\omega,-{\bf q}}$. Then we follow the standard procedure of the derivation of phonon-mediated electron-electron interaction \cite{SM}. The full action is $S=S_e+S_{ph}+S_{e-ph}$. We expand the partition function $\mathcal{Z}$ in $S_{e-ph}$, average over phonon degrees of freedom, reexponentiate it, obtaining the action for the electron-electron interaction $S_{e-e}=i\langle S_{e-ph}S_{e-ph}\rangle_{ph}/2$. The electron-phonon interaction action is
	\begin{eqnarray}
	S_{e-ph}=\int d\omega d{\bf q}F_{\omega,{\bf q}}\sum_{j=1,2}(a_{j,\omega,{\bf q}}+\bar{a}_{j,-\omega,-{\bf q}})\rho_{\omega,{\bf q}},\ \ \ \ 
	\end{eqnarray}
where the density of electrons is $\rho_{\omega,{\bf q}}=\int d{\bf K}d\Xi\bar{\psi}_{\omega+\Xi,{\bf K}+{\bf q}}\psi_{\Xi,\bf K}$ with $\bar{\psi}$ and $\psi$ being electron Grassmann fields; $\bar{a}$ and $a$ are phonon bosonic fields; the function $F_{\omega,{\bf q}}$ is the prefactor that usually contains material characteristics and constants, e.g. elastic constants; index $j$ denotes phonon bands 1 or 2.
	
After averaging over phonon degrees of freedom in $S_{e-e}$, we obtain the electron-electron interaction potential \cite{SM}:
\begin{eqnarray}
\nonumber V(\omega,{\bf q})=-\frac{F_{\omega,{\bf q}}F_{-\omega,-{\bf q}}}{2}\left[\frac{\omega_{1,{\bf q}}+\omega_{2,{\bf q}}-2\omega-2\delta_{\omega,{\bf q}}}{\delta_{\omega,{\bf q}}^2-(\omega-\omega_{1,{\bf q}})(\omega-\omega_{2,{\bf q}})}+\right.\\ \left.+\frac{\omega_{1,-{\bf q}}+\omega_{2,-{\bf q}}+2\omega-2\delta_{-\omega,-{\bf q}}}{\delta_{-\omega,-{\bf q}}^2-(\omega+\omega_{1,-{\bf q}})(\omega+\omega_{2,-{\bf q}})}\right].\ \ \ \ \ \ \ \ 
\end{eqnarray}
We assume that $\omega_{1,\pm{\bf q}},\omega_{2,\pm{\bf q}}\ll\delta_{\omega,{\bf q}},\delta_{-\omega,-{\bf q}},\omega$. This can be valid, for example, for acoustics phonons at small ${\bf q}$. Moreover, the phonon-phonon interaction can be enhanced externally e.g. by doping \cite{hong:jacs19} or laser pulsing \cite{zijlstra:prb06}. If we further assume that $\delta_{\pm\omega,{\bf q}}\gg \omega\gg\delta_{\pm\omega,-{\bf q}}$, we obtain
\begin{eqnarray}
V(\omega,{\bf q})\simeq -V(\omega,-{\bf q})\simeq \frac{F_{\omega,{\bf q}}F_{-\omega,-{\bf q}}}{\omega}.
\end{eqnarray}
Importantly, this electron-electron interaction potential is odd in momentum.

We also note that the periodic modulation of elastic properties e.g. due to spatiotemporal modulation should be incorporated into $F_{\omega,{\bf q}}$, because it is usually proportional to the elastic constants. Thus, $F_{\omega,{\bf q}}$ will have a peak around the modulation frequency which means that we can consider $\omega$ only around that frequency and the limit $\delta_{\pm\omega,-{\bf q}}\ll \omega\ll \delta_{\pm\omega,{\bf q}}$ is justified. In this model, we assume that spatiotemporal modulation generates non-reciprocity, but do not take into account possible non-equilibrium processes for electrons and phonons for simplicity.

{\it Multiband systems.--} Our formalism can be expanded to the case of multiband non-Hermitian superconductors, which we briefly discuss below. If electrons are photoemitted from several bands, the operators $c$ obtain additional band indices, and the anomalous Green's function is defined as $F_{\bar{\sigma},\alpha;\sigma,\beta}^{-{\bf p}, {\bf p}}(\tau_1,\tau_2)=\langle T [c_{-{\bf p},\bar{\sigma},\alpha}(\tau_1)c_{{\bf p}, \sigma,\beta}(\tau_2)]\rangle_0$ with $\alpha$ and $\beta$ denoting bands. Then, $I^{(2)}_{{\bf p},\uparrow;-{\bf p},\downarrow}$ can contain inter- and intra-band terms in case of several pulses addressing different bands \cite{kornich:prr21}.
If there is no complex spin structure of the Hamiltonian, which can be affected by the operator $\mathcal{T}$, the condition for a multiband Hamiltonian to be $\mathcal{PT}$-symmetric remains $\mathcal{H}_{\bf k}=\mathcal{H}^*_{\bf k}$. Then, the mean fields and other terms of the Hamiltonian must be real similarly to the single-band case. In this case, if the following condition applies: 
\begin{eqnarray}
F_{\bar{\sigma}\alpha;\sigma,\beta}^{-{\bf k},{\bf k}}(t_\alpha,t_\beta)=-\bar{F}_{\sigma,\beta;\bar{\sigma},\alpha}^{{\bf k},-{\bf k}}(t_\beta,t_\alpha), 
\end{eqnarray}
where $t_{\alpha,\beta}$ denote times of delta-shaped pulses addressing bands $\alpha$ and $\beta$, respectively, we again obtain a negative signal as in Eq. (\ref{eq:NegativeFl}).  

If the superconductor has a more complex Hamiltonian, the condition for being $\mathcal{PT}$-symmetric could be more complex. Then, we may obtain imaginary terms in the Hamiltonian which most likely lead to complex values of the Green's functions $F_{\bar{\sigma},\alpha;\sigma,\beta}^{-{\bf k},{\bf k}}(\omega)$ and $\bar{F}_{\sigma,\beta;\bar{\sigma},\alpha}^{{\bf k},-{\bf k}}(\omega)$. However, even then, if $(F_{\bar{\sigma}, \alpha; \sigma,\beta}^{-{\bf k},{\bf k}}(t_\alpha, t_\beta))^\dagger=-\bar{F}_{\sigma,\beta;\bar{\sigma},\alpha}^{{\bf k},-{\bf k}}(t_\beta, t_\alpha)$, we obtain a negative signal for the delta-shaped pulses in analogy to Eq. (\ref{eq:NegativeFl}). 

We mention in passing that in case of a more complicated expression for the Hamiltonian than Eq. (\ref{eq:HamiltonianSC}), the expression for the signal in Eq. (\ref{eq:DeltaIFull}) can also contain terms of the normal part of the Green's function because the two-point Green's function $G_{\sigma_2,\sigma_1,\sigma_1',\sigma_2'}^{{\bf k}_2,{\bf k}_1,{\bf k}_1',{\bf k}_2'}(\tau_2,\tau_1,\tau_1',\tau_2')$ contains different correlators and in general they may not all cancel with the correlators stemming from $I_{{\bf p},\sigma}^{(1)}I_{{\bf p}',\sigma'}^{(1)}$. If this is the case, the signal may be negative or positive depending on how the additional terms relate to the term with anomalous Green's functions.

In general, the signal from a non-Hermitian superconductor of the type of Eq. (\ref{eq:HamiltonianSC}) (not necessarily $\mathcal{PT}$-symmetric) can be positive or negative, because in the non-Hermitian case the signal $\Delta I_{{\bf p},\uparrow;-{\bf p},\downarrow}$ does not convert to the modulus square of an integral over one Green's function as in the Hermitian case \cite{stahl:prb19}. 

{\it Summary.--} In conclusion, we have analyzed angle-resolved photoelectron fluctuation spectroscopy (ARPFS) for a $\mathcal{PT}$-symmetric non-Hermitian superconductor. We have presented the non-Hermitian formalism for ARPFS and shown that the signal has to be negative in the single-band case. The negative fluctuations are a consequence of the asymmetry of electron-electron interaction in such $\mathcal{PT}$-symmetric superconductors, leading to attraction of electrons and repulsion of holes and vice versa. 
\begin{acknowledgments}
We acknowledge useful discussions with Sang-Jun Choi. This work was supported by the DFG (SPP1666 and SFB1170 ``ToCoTronics''), the W{\"u}rzburg-Dresden Cluster
of Excellence ct.qmat, EXC2147, project-id 390858490, and the Elitenetzwerk Bayern Graduate School on ``Topological Insulators''. Additionally, we acknowledge support from the High Tech Agenda Bavaria.
\end{acknowledgments}
\begin{widetext}
\section*{Supplemental Material }

\maketitle
\renewcommand{\theequation}{S\arabic{equation}}
\setcounter{equation}{0}
\renewcommand{\thefigure}{S\arabic{figure}}
\renewcommand{\figurename}{Supplementary Fig.}

\setcounter{figure}{0}
\renewcommand{\thesection}{S\arabic{section}}
\setcounter{section}{0}

In this Supplemental Material, we present additional details and calculations regarding: 1) derivation of the evolution operator for the left and right state vectors in the non-Hermitian case; 2) Fourier transformation of the anomalous Green's functions; 3) electron-electron interaction mediated by phonons from two bands with strong asymmetric interaction. 

\section{S1. Derivation of evolution operators for left and right state vectors in non-Hermitian case}
In this section, we derive the evolution operators for the left and right state vectors. As we consider the non-Hermitian case, it is not obvious how they look like.

The $\eta$-pseudo-Hermiticity is defined as \cite{mostafazadeh:jmp02}
\begin{eqnarray}
\label{eq:etaH}
H^\dagger=\eta H\eta^{-1},
\end{eqnarray}
where $\eta$ is a Hermitian and invertible linear operator. Then the $\eta$-product is defined as
\begin{eqnarray}
\langle\Phi|\Phi'\rangle_\eta=\langle\Phi|\eta|\Phi'\rangle.
\end{eqnarray}
The $\eta$-product can be derived from the definition of pseudo-Hermiticity and Schr\"odinger equation as follows. The Schr\"odinger equation for the right state vector is
\begin{eqnarray}
i\frac{\partial}{\partial t}|\Phi_R\rangle=H|\Phi_R\rangle.
\end{eqnarray}
If we take Hermitian conjugation of it and substitute the definition of $H^\dagger$ from Eq. (\ref{eq:etaH}), we obtain
\begin{eqnarray}
-i\frac{\partial}{\partial t}\langle\Phi_R|\eta=\langle\Phi_R|\eta H,
\end{eqnarray}
that gives
\begin{eqnarray}
\label{eq:SchroedPhiL}
-i\frac{\partial}{\partial t}\langle\Phi_L|=\langle\Phi_L|H.
\end{eqnarray}
Now let's derive the evolution operator in the interaction representation. For that, we represent the Hamiltonian in free and perturbation parts, $H=H_0+V$. Then, we use the Schr\"odinger equation following the standard procedure:
\begin{eqnarray}
i\frac{\partial}{\partial t}[e^{-iH_0t}|\Phi^I_R\rangle]=i[-iH_0e^{-iH_0t}|\Phi^I_R\rangle+e^{-iH_0t}\partial_t|\Phi^I_R\rangle ]=(H_0+V)|\Phi_R\rangle.
\end{eqnarray}
The terms with $H_0$ cancel each other and we obtain
\begin{eqnarray}
i\partial_t|\Phi^I_R\rangle =e^{iH_0t}Ve^{-iH_0t}e^{iH_0t}|\Phi_R\rangle=V^I|\Phi^I_R\rangle.
\end{eqnarray}
This is a standard expression of the Schr\"odinger equation in the interaction picture. We can do exactly the same with Eq. (\ref{eq:SchroedPhiL}) with the only change that $\langle\Phi_L^I|=\langle\Phi_L|e^{-iH_0t}$ and obtain
\begin{eqnarray}
-i\frac{\partial}{\partial t}\langle\Phi_L^I|=\langle\Phi_L|e^{-iH_0t}e^{iH_0t} Ve^{-iH_0t}=\langle\Phi_L^I|V^I.
\end{eqnarray}
the evolution operator for $\langle \Phi_L^I|$ is
\begin{eqnarray}
\mathcal{S}_L=\bar{T}e^{i\int_0^td t'V^I(t')},
\end{eqnarray}
and for $|\Phi_R^I\rangle$ is
\begin{eqnarray}
\mathcal{S}_R=Te^{-i\int_0^td t'V^I(t')}.
\end{eqnarray}

\section{S2. Fourier transformation of the anomalous Green's functions}
The poles of the Green's functions derived in Eq. (8) of the main text are $\omega_{1,2}=\pm\sqrt{\varepsilon_k^2-\Delta_{\bf k}^2}$. We perform the corresponding integration following the standard procedure and firstly add $\pm i\eta$ to the denominators of the Green's functions with $\eta$ being infinitesimally small, $\eta\rightarrow+0$. As the electron Green's function usually has $i\eta$ and the hole Green's function has $-i\eta$, we put $i\eta$ to the denominator of $\bar{F}$ and $-i\eta$ to the denominator of $F$. If we express denominators in terms of products and expand the square roots in $\eta$, we obtain
\begin{eqnarray}
	\bar{F}_{\uparrow,\downarrow}^{{\bf k},-{\bf k}}(\omega)&=&\frac{\Delta_{\bf k}}{[\omega-\sqrt{\varepsilon_k^2-\Delta_{\bf k}^2}+i\eta][\omega+\sqrt{\varepsilon_k^2-\Delta_{\bf k}^2}-i\eta]}, \\
	F_{\downarrow,\uparrow}^{-{\bf k},{\bf k}}(\omega)&=&\frac{-\Delta_{\bf k}}{[\omega-\sqrt{\varepsilon_k^2-\Delta_{\bf k}^2}-i\eta][\omega+\sqrt{\varepsilon_k^2-\Delta_{\bf k}^2}+i\eta]}.
\end{eqnarray}
This implies that $\bar{F}_{\uparrow,\downarrow}^{{\bf k},-{\bf k}}$ has a pole $\omega_2+i\eta$ and $F_{\downarrow,\uparrow}^{-{\bf k},{\bf k}}$ has a pole $\omega_1+i\eta$ in the upper half plane. Building a contour of half-circle shape with the radius $R\rightarrow\infty$ and employing residues (see e.g. Ref. \onlinecite{kornich:prr21}), we obtain for the anomalous Green's functions in the time domain
\begin{eqnarray}
\bar{F}_{\uparrow,\downarrow}^{{\bf k},-{\bf k}}(t)&=&-\frac{i\Delta_{\bf k}e^{-it\sqrt{\varepsilon_k^2-\Delta_{\bf k}^2}}}{2\sqrt{\varepsilon_k^2-\Delta_{\bf k}^2}},\\
F_{\downarrow,\uparrow}^{-{\bf k},{\bf k}}(t)&=&-\frac{i\Delta_{\bf k}e^{it\sqrt{\varepsilon_k^2-\Delta_{\bf k}^2}}}{2\sqrt{\varepsilon_k^2-\Delta_{\bf k}^2}}.
\end{eqnarray}
We can see that $\bar{F}_{\uparrow,\downarrow}^{{\bf k},-{\bf k}}(t)\neq (F_{\downarrow,\uparrow}^{-{\bf k},{\bf k}}(t))^\dagger$.

\section{S3. Derivation of Electron-electron interaction mediated by phonons from two interacting bands}
Let's consider electrons interacting with phonons. Then, the partition function is given by
\begin{eqnarray}
\mathcal{Z}=\int \mathcal{D}[\bar{\psi},\psi,\bar{a},a]\exp{[i(S_{ph}+S_e+S_{e-ph})]},
\end{eqnarray}
where $S_e$ is the action of electrons with $\psi$ and $\bar{\psi}$ being Grassmann fields of electrons and $S_{ph}$ is the action of phonons with $\bar{a}$ and $a$ being bosonic fields of phonons. We make a common assumption that $\langle a\rangle=\langle\bar{a}\rangle=0$. The action $S_{e-ph}$ describes electron-phonon interaction.

Let's consider $S_{e-ph}$ as a perturbation, and expand the exponent in it. Then, we average over phononic degrees of freedom assuming that $S_{e-ph}$ is linear in $a$ and $\bar{a}$, and thus all odd powers of $S_{e-ph}$ are zero. In such a way, we obtain
\begin{eqnarray}
\mathcal{Z}=\int \mathcal{D}[\bar{\psi},\psi,\bar{a},a]\exp{[i(S_{ph}+S_e)]}(1+iS_{e-ph}-\frac{1}{2}S_{e-ph}S_{e-ph}+...)=\int \mathcal{D}[\bar{\psi},\psi]\exp{[iS_e-\frac{1}{2}\langle S_{e-ph}S_{e-ph}\rangle]}.\ \ \
\end{eqnarray}
The last term is the action for electron-electron interaction mediated by phonons $S_{e-e}=i\langle S_{e-ph}S_{e-ph}\rangle/2$.

Now we consider the phonon action in detail. We assume that there are two phonon bands $\omega_{1,{\bf q}}$ and $\omega_{2,{\bf q}}$ that interact with the off-diagonal element $\delta_{\omega,{\bf q}}$:
\begin{eqnarray}
S_{ph}=\int \begin{pmatrix}\bar{a}_{1,\omega,{\bf q}} & \bar{a}_{2,\omega,{\bf q}}\end{pmatrix}\begin{pmatrix}\omega-\omega_{1,{\bf q}} & -\delta_{\omega,{\bf q}}\\ -\delta_{\omega,{\bf q}} & \omega-\omega_{2,{\bf q}}\end{pmatrix}\begin{pmatrix}a_{1,\omega,{\bf q}} \\ a_{2,\omega,{\bf q}}\end{pmatrix}d\omega d{\bf q}.
\end{eqnarray}
The electron-phonon interaction is described as
\begin{eqnarray}
S_{e-ph}=\int d\omega d{\bf q}F_{\omega,{\bf q}}\sum_{j=1,2}(a_{j,\omega,{\bf q}}+\bar{a}_{j,-\omega,-{\bf q}})\int d\Xi d{\bf K}\bar{\psi}_{\omega+\Xi,{\bf K}+{\bf q}}\psi_{\Xi,{\bf K}}=\int d\omega d{\bf q}F_{\omega,{\bf q}}\sum_{j=1,2}(a_{j,\omega,{\bf q}}+\bar{a}_{j,-\omega,-{\bf q}})\rho_{\omega,{\bf q}}.\ \ \ \ \ \ \ \ \ 
\end{eqnarray}
Here, $F_{\omega,{\bf q}}$ is the prefactor that usually contains material characteristics and constants, for instance, elastic constants. This gives
\begin{eqnarray}
\langle S_{e-ph}S_{e-ph}\rangle=\int \mathcal{D}[\bar{a},a]e^{iS_{ph}}S_{e-ph}S_{e-ph}=\\ \nonumber=\int d{\bf q}d{\bf q'}d\omega d\omega'F_{\omega,{\bf q}}F_{\omega',{\bf q'}}\rho_{\omega,{\bf q}}\rho_{\omega',{\bf q'}}\int \mathcal{D}[\bar{a},a]\sum_{i,j=1,2}[a_{i,\omega,{\bf q}}\bar{a}_{j,-\omega',-{\bf q}'}+\bar{a}_{i,-\omega,-{\bf q}}a_{j,\omega',{\bf q}'}] e^{iS_{ph}}.
\end{eqnarray}
Now we will use the property that
\begin{eqnarray}
\langle\bar{a}_{\bf q_1}a_{\bf q_2}\rangle=\int \mathcal{D}[\bar{a},a]\bar{a}_{\bf q_1}a_{\bf q_2}e^{-\sum_q\bar{a}_{\bf q}A_{\bf q}a_{\bf q}}=\frac{\delta_{{\bf q}_1,{\bf q}_2}}{A_{{\bf q}_1}},
\end{eqnarray}
in order to perform averaging. After averaging, we obtain
\begin{eqnarray}
\langle S_{e-ph}S_{e-ph}\rangle&=&i\int d{\bf q}d\omega F_{\omega,{\bf q}}F_{-\omega,-{\bf q}}\rho_{\omega,\bf q}\rho_{-\omega,-{\bf q}}\times \\ \nonumber&&\left[\frac{\omega_{1,{\bf q}}+\omega_{2,{\bf q}}-2\omega-2\delta_{\omega,{\bf q}}}{\delta_{\omega,{\bf q}}^2-(\omega-\omega_{1,{\bf q}})(\omega-\omega_{2,{\bf q}})}+\frac{\omega_{1,-{\bf q}}+\omega_{2,-{\bf q}}+2\omega-2\delta_{-\omega,-{\bf q}}}{\delta_{-\omega,-{\bf q}}^2-(\omega+\omega_{1,-{\bf q}})(\omega+\omega_{2,-{\bf q}})}\right].\ \ \ 
\end{eqnarray}
This yields the expression for the electron-electron interaction potential that we state in the main text in Eq. (17):
\begin{eqnarray}
V(\omega,{\bf q})=-\frac{F_{\omega,{\bf q}}F_{-\omega,-{\bf q}}}{2}\left[\frac{\omega_{1,{\bf q}}+\omega_{2,{\bf q}}-2\omega-2\delta_{\omega,{\bf q}}}{\delta_{\omega,{\bf q}}^2-(\omega-\omega_{1,{\bf q}})(\omega-\omega_{2,{\bf q}})}+\frac{\omega_{1,-{\bf q}}+\omega_{2,-{\bf q}}+2\omega-2\delta_{-\omega,-{\bf q}}}{\delta_{-\omega,-{\bf q}}^2-(\omega+\omega_{1,-{\bf q}})(\omega+\omega_{2,-{\bf q}})}\right].
\end{eqnarray}
Now let's assume that $\omega_{1,\pm{\bf q}},\omega_{2,\pm{\bf q}}\ll \delta_{\omega,{\bf q}},\delta_{-\omega,-{\bf q}},\omega$. Then we obtain
\begin{eqnarray}
V(\omega,{\bf q})\simeq F_{\omega,{\bf q}}F_{-\omega,-{\bf q}}\left[\frac{1}{\delta_{\omega,{\bf q}}-\omega}+\frac{1}{\delta_{-\omega,-{\bf q}}+\omega}\right]=F_{\omega,{\bf q}}F_{-\omega,-{\bf q}}\left[\frac{\delta_{-\omega,-{\bf q}}+\delta_{\omega,{\bf q}}}{(\delta_{\omega,{\bf q}}-\omega)(\delta_{-\omega,-{\bf q}}+\omega)}\right].
\end{eqnarray}
If the phonon interaction is strongly asymmetric in momentum space, $\delta_{\pm\omega,{\bf q}},\omega\gg\delta_{\pm\omega,-{\bf q}}$, then
\begin{eqnarray}
V(\omega,{\bf q})&\simeq& F_{\omega,{\bf q}}F_{-\omega,-{\bf q}}\frac{\delta_{\omega,{\bf q}}}{\omega(\delta_{\omega,{\bf q}}-\omega)},\\
V(\omega,-{\bf q})&\simeq& F_{\omega,{\bf q}}F_{-\omega,-{\bf q}}\frac{\delta_{-\omega,{\bf q}}}{-\omega(\delta_{-\omega,{\bf q}}+\omega)}.
\end{eqnarray}
If we further assume, $\delta_{\pm\omega,{\bf q}}\gg \omega$, we obtain Eq. (18) from the main text,
\begin{eqnarray}
V(\omega,{\bf q})\simeq -V(\omega,-{\bf q})\simeq\frac{F_{\omega,{\bf q}}F_{-\omega,-{\bf q}}}{\omega}.
\end{eqnarray}
Thus we obtain odd or asymmetric electron-electron interaction potential due to an externally-induced asymmetry in the phonon system, namely, the phonon-phonon interaction. The band structure of phonons, $\omega_{1,{\bf q}}$ and $\omega_{2,{\bf q}}$, can also be asymmetric, but here we have not used this property.

\end{widetext}

\end{document}